\documentclass[conference]{IEEEtran}
\IEEEoverridecommandlockouts

\usepackage{listings}
\usepackage{xcolor}
\usepackage{xspace}
\usepackage{hyperref}
\usepackage{multirow}
\usepackage{makecell}
\usepackage[normalem]{ulem}
\usepackage{enumitem}
\usepackage[most]{tcolorbox}
\usepackage{cite}
\usepackage{amsmath,amssymb,amsfonts}
\usepackage{algorithmic}
\usepackage{graphicx}
\usepackage{textcomp}
\usepackage{xcolor}
\usepackage{caption}
\def\BibTeX{{\rm B\kern-.05em{\sc i\kern-.025em b}\kern-.08em
    T\kern-.1667em\lower.7ex\hbox{E}\kern-.125emX}}

\newtcolorbox{rqbox}{
  colback=white,
  colframe=black,
  boxrule=1pt,
  arc=8pt,
  left=8pt,
  right=8pt,
  top=8pt,
  bottom=8pt
}

\newcommand{\ie}{\emph{i.e.}\xspace}

\newcommand{\eg}{\emph{e.g.}\xspace}
\newcommand{\fig}{Fig.\xspace}
\newcommand{\tab}{Table\xspace}

\definecolor{light-gray}{gray}{0.87}
\newcommand{\lightgray}[1]{\colorbox{light-gray}{#1}}
\newcommand{\lightyellow}[1]{\colorbox{yellow!25}{#1}}
\newcommand{\lightred}[1]{\colorbox{red!25}{#1}}

\newcommand{\sepAnd}{\&*\&}

\begin{document}
\bstctlcite{IEEEexample:BSTcontrol}

\title{An Empirical Study of LLM-Generated Specifications for VeriFast
}

\author{\IEEEauthorblockN{Wen Fan}
\IEEEauthorblockA{\textit{Purdue University} \\
West Lafayette, Indiana, USA \\
fan372@purdue.edu}
\and
\IEEEauthorblockN{Minh Tran}
\IEEEauthorblockA{\textit{Purdue University} \\
West Lafayette, Indiana, USA \\
tran299@purdue.edu}
\and
\IEEEauthorblockN{Sanya Dod}
\IEEEauthorblockA{\textit{Purdue University} \\
West Lafayette, Indiana, USA \\
sdod@purdue.edu}
\and
\IEEEauthorblockN{Xin Hu}
\IEEEauthorblockA{\textit{University of Michigan} \\
Ann Arbor, Michigan, USA \\
hsinhu@umich.edu}
\and
\IEEEauthorblockN{Marilyn Rego}
\IEEEauthorblockA{\textit{University of Michigan} \\
Ann Arbor, Michigan, USA \\
mrego@umich.edu}
\and
\IEEEauthorblockN{Danning Xie}
\IEEEauthorblockA{\textit{Meta} \\
Menlo Park, California, USA \\
xiedn1105@gmail.com}
\and
\IEEEauthorblockN{Jenna DiVincenzo}
\IEEEauthorblockA{\textit{Purdue University} \\
West Lafayette, Indiana, USA \\
jennad@purdue.edu}
\and
\IEEEauthorblockN{Lin Tan}
\IEEEauthorblockA{\textit{Purdue University} \\
West Lafayette, Indiana, USA \\
lintan@purdue.edu}
}

\maketitle

\begin{abstract}
Static verification tools can assure industrial scale software, but require significant human labor to write specifications. This is particularly true of static verifiers based on separation logic (SL verifiers), which excel at verifying heap-manipulating programs, but require many complex auxiliary specifications to reason about heap structure.
Recent work applies large language models (LLMs) to generate code, tests, and proofs, including specifications for verifiers, but mostly targeting non-SL verifiers.
To address this gap, this paper thoroughly evaluates how well LLMs perform when prompted to generate specifications for verifying 303 C functions with the SL verifier VeriFast. We explored eight prompting approaches, ten LLMs, and three input types in two stages. Quantitative and qualitative analyses are used to assess the LLM-generated code and specifications for functional behavior, verifiability and errors. 
The results show that LLMs preserve functional behavior in source code and specifications (both over 91\%), but achieve modest verification success (31.4\%). Using Gemini 2.5 Pro and providing formal contracts lead to higher success rates in our setting.
Moreover, most errors (94\%) come from LLMs' mistakes in the domain-specific knowledge of SL verifiers such as VeriFast. %
These findings provide guidance for optimizing LLM-generated specifications for SL verifiers.
\end{abstract}


\section{Introduction}
Advancements in \emph{satisfiability modulo theories (SMT) solvers} \cite{moskewicz2001chaff} have driven the development of automated deductive verification tools---also known as auto-active, static, Hoare-logic \cite{hoare1969axiomatic}, or design-by-contract verifiers. Such verifiers work by taking specifications written in a first-order mathematical logic on program components (\eg pre- and postconditions on functions), which denote the intended program behavior, and then statically analyzing the program to prove its adherence to the specifications. The analysis builds up proof obligations (\eg at the end of a function the analysis's data should imply the postcondition) that are sent to an SMT solver to discharge. Successful verification of a program means that it will not violate its specifications at run time. Mature static verifiers, such as Dafny \cite{leino2010dafny} and VeriFast \cite{lattuada2023verus}, have been successfully applied in industry. Dafny was used to verify the functional correctness of Amazon Web Services' (AWS) authorization engine \cite{chakarov2025formally}, and VeriFast was used to verify the absence of safety errors (\eg dividing by zero or illegal memory accesses) in two smart card applets, a Linux device driver, and an embedded Linux network management component \cite{philippaerts2014software}.

While promising, static verifiers have not achieved the ideal in automation where users only need to specify their intended behavior for code. Instead, verifiers require auxiliary specifications, such as loop invariants, inductive lemmas, and folds and unfolds of predicates to make up for limitations in proof automation. Furthermore, writing these specifications requires internal knowledge of the verifier's analysis and its status throughout the verification process. It should not come as a surprise then that it took 4 expert person-years to verify the AWS authorization engine \cite{chakarov2025formally}.
In response, and with the success of large language models (LLMs) at generating code \cite{chen2022codet,sarsa2022automatic}, test-cases \cite{wang2024software,xia2024fuzz4all}, program invariants \cite{ernst2007daikon,pei2023can,xie2023impact,sun2025classinvgen, cao2025clause2inv}, and proofs in proof assistants \cite{zheng2023lyra,yang2024leandojo, jiang2021lisa,welleck2023llmstep,first2023baldur}, there has been a surge of recent work (all in the last 3 years) using LLMs to successfully generate specifications for such verifiers \cite{sevenhuijsen2025vecogen, wen2024enchanting, janssen2024can, kamath2024leveraging, ma2025specgen, misu2024towards, mugnier2025laurel, pascoal2025automatic, silva2024leveraging, lahiri2024evaluating, poesia2024dafny, yao2023leveraging, chen2025automated, yang2025autoverus, rego2025evaluating, banerjee2026dafnypro, silva2025inferring}. 

Only one of these works \cite{rego2025evaluating} concerns static verifiers based on \emph{separation logic} \cite{reynolds2002separation, parkinson2005separation} (SL verifiers), such as Viper \cite{muller2016viper}, Gillian \cite{fragoso2020gillian}, and VeriFast \cite{jacobs2011verifast}.

This understudy is unfortunate %
as SL verifiers in particular excel at verifying both the functional and safety properties of heap-manipulating programs at scale \cite{philippaerts2014software}, but they suffer from requiring auxiliary specifications that are further complicated by reasoning about the heap. This leads to specification patterns that are unique to SL verifiers, \eg numerous folds and unfolds, loop invariants, and lemmas breaking down or connecting predicates describing parts of the heap, and are worth studying as a result. %

Therefore, this paper seeks to empirically and thoroughly evaluate how well existing LLMs perform when prompted to generate specifications for verifying C programs by the SL verifier VeriFast \cite{jacobs2011verifast}. 
In particular, we examine both their ability to preserve 
functional behavior and produce verifiable specifications across 
settings, and analyze the errors that arise when verification fails.

To conduct the study, we 
develop a dataset of 303 functions from 60 C programs with three different input versions of each function containing their functional behavior specified in natural language or VeriFast's specification language. Our dataset also includes fully specified and verified (ground truth) versions of these functions. Throughout our study, LLMs are prompted to generate missing specifications needed to verify a function given its input files; the resulting output files are then quantitatively and qualitatively analyzed.
In particular, our study proceeds in two stages: 1) we conduct pilot studies on 45 functions to select the best performing prompting approach out of eight and three best performing LLMs out of ten, and then 2) we conduct a full study on the full dataset with the selected prompt approach and LLMs. For the pilot studies, we make selections based on verification success rates, the number of verification errors and their severity, and the number of functional behavior changes made to function pre- and postconditions and source code. In the full study, we manually code source code and pre- and postconditions in the output files to assess their functional behavior. We also calculate the percentage of functions passing verification with VeriFast. Finally, we sample 50 failing functions and manually record and fix errors until the function is verified. During this process, we record the error severity and the fix applied.

We find that LLMs largely preserve functional behavior in both source code (91.0\%) and pre- and postconditions (92\%), but have a fairly modest verification success rate at 31.4\%. Moreover, we find Gemini 2.5 Pro and providing stronger pre- and postconditions in the input achieve the best performance for verifiability. Lastly, our error analysis shows that the majority of errors (94\%) arise from limitations in the knowledge of SL verifier (\eg syntax and heap reasoning), rather than general logical reasoning. We also provide actionable insights from our results, such as the suggestion to provide better heap related debugging information from a SL verifier to LLMs when verification fails %
and enabling the auto-feature of VeriFast--both to reduce verification errors and improve success rates.

In summary, this paper makes the following contributions:
\begin{itemize}
    \item The first empirical and thorough study of LLMs' ability to generate specifications for program verification with an SL verifier using prompt engineering. In particular, we consider 8 different prompting approaches, 10 different LLMs, and 3 different input types.
    \item A dataset of 60 C programs with 303 functions that each have three versions with specifications describing their functional behavior in different formats. The dataset also includes a fourth fully specified and verified version of each program and function. 
    \item Analysis results and actionable insights (\S\ref{sec:results}) that provide a baseline and guidance for building LLM-based verification environments for proof synthesis in SL verifiers, particularly VeriFast.
    \item 354 LLM specified C functions failing to verify with VeriFast along with human coded errors for each function that capture their severity, location, and fix/root cause.
\end{itemize}

\section{Background \& Motivation}

\subsection{The Benefits of Separation Logic Based Static Verifiers}
Static verifiers verify programs by checking that each function adheres to its intended behavior written as pre- and postconditions in a mathematical logic. For heap-manipulating functions, this logic is a first-order logic extended with \emph{separation logic} (SL) \cite{reynolds2002separation} or its variants, such as \emph{implicit dynamic frames} (IDF) \cite{smans2012implicit}. SL allows one to express how the shape of the heap or its parts change during execution; and, static verifiers supporting SL (\ie SL verifiers) ensure the function respects this behavior.
For example, \fig \ref{fig:ll-appendtail} contains a simple C implementation of an \texttt{append\_tail} function for singly-linked lists, which inserts a new node with a given value at the tail end of a given list. It also contains specifications highlighted in \lightgray{gray} leading to the successful verification of \texttt{append\_tail} with the SL verifier VeriFast \cite{jacobs2011verifast}.
\begin{figure*}[t]
\input{figures/ll-appendtail}
\caption{Static verification in VeriFast of the \texttt{append\_tail} function for singly-linked lists.}
\label{fig:ll-appendtail}
\end{figure*}
The precondition \texttt{llist(head)} (line \ref{ll-pre}) specifies the input list starting at \texttt{head} must be acyclic, and the postcondition \texttt{llist(result)} (line \ref{ll-post}) similarly denotes the returned list must be acyclic. In particular, the recursive predicate \texttt{lseg} (lines \ref{ll-lseg-start}-\ref{ll-mallocblock}), which defines \texttt{llist} (line \ref{ll-llist}), uses \emph{separation logic arrow} (\ie the points-to assertion or singleton heap) to express heap locations and the values they hold, \eg \texttt{from->val |-> ?val} states the heap location \texttt{from->val} holds the value \texttt{val}. The points-to assertion also denotes unique ownership of the given heap location, \eg \texttt{from->val}. The \texttt{lseg} predicate also relies on the \emph{separating conjunction} \texttt{\sepAnd} to denote disjointness in memory. For example, \texttt{y->next |-> ?ynext \sepAnd~ x->next |-> ?xnext} implies \texttt{y->next} and \texttt{x->next} are distinct heap locations, \ie \texttt{y != x}. Then altogether, \texttt{lseg} specifies that all the heap locations for nodes in a list segment from the \texttt{from} node to the \texttt{to} node are separate from one another in memory, \ie that the list segment is acyclic. It also denotes ownership of these heap locations. As a result, \texttt{append\_tail}'s pre- and postconditions state: \texttt{append\_tail} only accesses the heap locations for the given acyclic list, and returns the heap locations for an acyclic list. 

By supporting SL, SL verifiers are able to assure heap-manipulating programs for functional properties involving the heap, such as preservation of list acyclicness.
Further, SL verifiers utilize the ownership and separation constraints from SL to guarantee the absence of access errors, such as dangling or null-pointer dereferences, and memory leaks \footnote{VeriFast uses malloc block (\eg at line \ref{ll-mallocblock}) to help detect memory leaks.}.

\subsection{The Specification Burden of  SL Verifiers}
\label{sec:sl-burden}
Unfortunately, SL verifiers not only require traditional auxiliary specifications (\eg predicates, loop invariants, and inductive lemmas), but also additional complexity from ownership tracking and checking unique to SL verifiers. Notice, in \fig \ref{fig:ll-appendtail}, verifying \texttt{append\_tail} requires 34 lines of auxiliary specification code compared to 16 lines of program code\footnote{Empty lines and trivial lines (\eg with only \{, \}, /*@ or @*/) are excluded.}.

In SL verifiers, \emph{predicates} enable the expression of heap properties for recursive heap data structures, such as \texttt{lseg} and \texttt{llist} on lines \ref{ll-lseg-start}-\ref{ll-llist} in \fig \ref{fig:ll-appendtail} for lists. But, SL verifiers cannot reliably automate unrolling recursive predicates; so instead, \emph{folds and unfolds} (\emph{opens and closes} in VeriFast) must be specified to control the availability of predicate information during verification, including for heap access checks\footnote{Although VeriFast can infer some heap chunks (\eg, \texttt{llist(head)} on line \ref{ll-close-llhead}) automatically, fold/unfold cannot be totally eliminated in SL verifiers.}. For example, \texttt{open llist(head)} on line \ref{ll-open-llhead} in \fig \ref{fig:ll-appendtail} tells VeriFast to consume the \texttt{list(head)} predicate from \texttt{append\_tail}'s precondition (line \ref{ll-pre}) and produce the predicate's body \texttt{lseg(head,0)} for use in the following open on line \ref{ll-open-lseghead0}. The second open provides the \texttt{head->next |-> ?next} points-to assertion, which justifies \texttt{curr->next} in the while loop's condition (line \ref{ll-loop-start}) on entry. Conversely, the \texttt{close llist(head)} on line \ref{ll-close-llhead} tells VeriFast to consume \texttt{list(head)}'s body \texttt{lseg(head,0)} and produce \texttt{list(head)} itself, which is used to prove \texttt{append\_tail}'s postcondition (line \ref{ll-post}) after returning \texttt{head} (line \ref{ll-return}).

\emph{Loop invariants}, which specify properties preserved across all iterations of a loop, enable static verifiers to verify all execution paths of a loop. They are 1) checked to hold before the loop, and 2) used to verify the loop body similar to a function body given the loop invariants as both the pre- and postconditions. In SL verifiers, loop invariants must additionally specify ownership of heap locations accessed by the loop to ensure safe access. Following these constraints, the loop invariant for the while loop in \fig \ref{fig:ll-appendtail} on lines \ref{ll-loopinv-start}-\ref{ll-loopinv-end} specifies that the current node is always non-null (line \ref{ll-loopinv-start}) and the footprint of the loop in three parts: the list segmented from its head to the current node (line \ref{ll-loopinv-start}), the points-to assertions for the current node (line \ref{ll-loopinv-curr}), and the list segmented from the node after the current one to the end (line \ref{ll-loopinv-end}). The points-to assertions and list segment after the current node provide access to heap locations used in the loop (\eg \texttt{curr->next} on lines \ref{ll-loop-start} and \ref{ll-loopbdy-currnext}). The list segment from the head to current node retains these heap locations for use after the loop.

Finally, SL verifiers need \emph{inductive lemmas} to break down or connect properties about the heap. For example, at the end of \texttt{append\_tail}, we need to leverage \texttt{lseg(curr,0)} (produced on line \ref{ll-close-llsegcurr0}) and \texttt{lseg(head,curr)} (provided by loop invariant) to obtain \texttt{lseg(head,0)} for the postcondition (line \ref{ll-post}). SL verifiers cannot automatically prove transitivity of list segments, so the inductive lemma \texttt{lseg\_merge} specified on lines \ref{ll-merge-start}-\ref{ll-merge-end} provides this proof. Then, \texttt{lseg\_merge} is called on line \ref{ll-merge} to help prove the postcondition. %

\subsection{Large Language Models (LLMs) \& Prompt Engineering}
LLMs are transformer-based models pretrained on large-scale natural language and code corpora to learn general linguistic and structural patterns~\cite{vaswani2017attention, touvron2023llama, liu2024deepseek, nijkamp2022codegen, lozhkov2024starcoder}. 
Rather than task-specific fine-tuning, LLMs can be adapted at inference time via in-context learning~\cite{brown2020language, chowdhery2023palm, grattafiori2024llama, xie2021explanation} with prompt engineering. 
Chain-of-Thought (CoT)~\cite{wei2022chain} and Retrieval-Augmented Generation (RAG)~\cite{lewis2020retrieval} are widely used techniques for improving LLM performance on tasks requiring structured reasoning and domain knowledge. CoT encourages explicit intermediate reasoning, which has been shown to improve accuracy on complex reasoning tasks~\cite{lewkowycz2022solving, zelikman2022star}. RAG augments the model input with task-relevant contextual information retrieved from an external corpus, allowing the model to leverage additional background knowledge beyond what is implicitly captured during pretraining~\cite{yang2024swe, zhang2023repocoder, luo2024repoagent}. These prompting techniques enable LLMs to better leverage domain-specific knowledge and reasoning patterns, leading to improved performance on specialized tasks requiring structured reasoning and expertise, such as test generation~\cite{wang2024software, zheng2025large, deng2023large, lemieux2023codamosa}, code generation~\cite{zhang2024autocoderover, jimenez2023swe, ding2024cocomic}, and proof assistants~\cite{thompson2025rango, lu2025adaptive, miku2024magnushammer, zheng2023lyra, first2023baldur}.

In RAG, the retrieved context is usually retrieval using similarity between the input query and candidate contexts. Sparse retrieval (RAG-sparse) relies on lexical similarity, such as term overlap measured by traditional information retrieval methods (e.g., TF-IDF and BM25~\cite{bm25}), while dense retrieval (RAG-dense)~\cite{karpukhin2020dense} selects context based on the similarity of neural embeddings, enabling retrieval of semantically related content even when surface forms differ.

\subsection{VeriFast}
VeriFast \cite{jacobs2011verifast, verifastrepo2025} is an SL verifier being actively developed by the imec-DistriNet research group at KU Leuven in Leuven, Belgium.
VeriFast supports the modular verification of single and multi-threaded C, Java, and unsafe Rust code for functional and memory safety properties.
These properties may be expressed with inductive datatypes, primitive recursive pure functions over these datatypes, first-order logic over data values, abstract predicates, SL points-to assertions and separating conjunctions. To aid in the verification of such rich specifications, users may write loop invariants, opens and closes of predicates, and lemmas \cite{jacobs2011verifast, verifastrepo2025} or rely on limited symbolic inference features where available \cite{vogels2011annotation}.
VeriFast verifies a program using a forward symbolic execution algorithm, which sends assertions over data values to an SMT solver to be checked against the current path condition \cite{jacobs2011verifast}. We use the default SMT solver supported by VeriFast, which is Z3 \cite{de2008z3}.

VeriFast was chosen due to its maturity and active development since 2011 (version 25.11 released on Nov. 27, 2025). It is supported on Windows, Linux, and macOS with easy to install binaries and has a VS Code extension. It also has detailed tutorials for C and Rust, detailed documentation, and Zulip chatroom support \cite{verifastrepo2025}. Finally, VeriFast's GitHub repository \cite{verifastrepo2025} contains 770 test C programs derived from real-world code to be considered for our test dataset (\S\ref{sec:test-dataset}). %

\section{Study Design}
\label{sec:study-design}

\begin{figure}[t]
\centering
\includegraphics[width=\columnwidth]{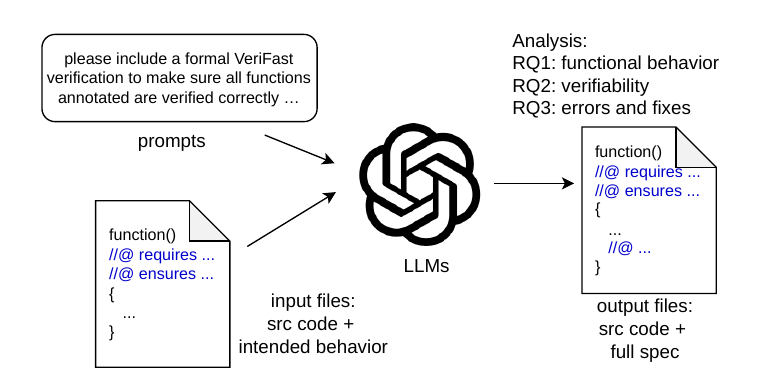}
\caption{The overview of the study}
\label{fig:overview}
\end{figure}

To evaluate LLMs on generating VeriFast specifications, we aim to answer the following research questions: 

\begin{itemize}
  \item[RQ1] How well do LLMs preserve \emph{functional behavior} when generating specifications for proofs of correctness in VeriFast across different prompts, input types, and LLMs?
  \item[RQ2] How successful are LLMs at generating specifications that result in \emph{verified} code with VeriFast across different prompts, input types, and LLMs?
  \item[RQ3] When LLMs generate specifications that fail to verify with VeriFast, what \emph{errors} are produced and how can they be fixed?
\end{itemize}
This section describes the empirical studies we perform to answer RQ1-3. %
First, we conduct two small-scale studies to pick the best prompting method out of 8 candidates and the best 3 LLMs out of 10 candidates for further study. These studies follow the same procedure as our larger-scale study just over the subset of our dataset (\S\ref{sec:test-dataset}) not affected by data leakage. 
In particular, for each chosen LLM, input program, and prompt, we feed the input program and prompt to the LLM and record the output program produced. Input programs contain functions specified with their intended behavior in 1 of 3 forms, and prompts contain instructions and details (depending on prompt type) that instruct the model to produce a verified version of the input with VeriFast. Then, we evaluate the quality of the output program quantitatively and qualitatively for its preservation of functional behavior, verifiability, and errors. This workflow is illustrated in \fig \ref{fig:overview}. 
The larger-scale study follows this workflow for the best prompt approach, best 3 LLMs, and all input types and programs in our dataset.

\subsection{Test Dataset}
\label{sec:test-dataset}
Our dataset includes four different versions of 303 functions from 60 C programs (\ie,  $\approx$ 5 functions per program), with 50 from the VeriFast GitHub repository (\S\ref{sec:dataset-verifast}) and 10 developed by  (\S\ref{sec:dataset-diy}). Three of the variants are only specified with functional behavior in different formats (\S\ref{sec:dataset-inputs}) and serve as input files to the LLMs for verification. The fourth variant is fully specified and verified with VeriFast and serves as a ground truth for our analyses. The programs in this fourth form have an average of 43 lines of source code (ranging from 3 to 143) and 81 lines of specifications (ranging from 5 to 600). Of the 303 functions, 45 contain concurrency, 39 loops, 134 need recursive predicates for verification, 
and 85 are other non-trivial heap manipulating functions \footnote{Note that only one of these tags is assigned to a function. Concurrency outranks loops, which outranks recursive predicates.}.

\subsubsection{50 Programs Derived from the VeriFast GitHub repository} \label{sec:dataset-verifast}
VeriFast's GitHub repository contains 770 C programs, from example, tutorial, test, and library folders, 
that are specified with VeriFast's specification language and represent real code. %
From the 770 programs, we started with the 160 single-file ones from the example and tutorial folders. Of the 160 files, 46 do not have code or SL specifications, 49 are duplicates or in the tutorial used in prompts, 10 had verification errors with VeriFast 24.8.30 that we were unable to fix, and 5 could not have inputs prepared. After dropping these files, we are left with 50 verified programs--with 258 functions, an ave. of 44 lines of code (rge: 3-143), and an ave. of 91 lines of specifications (rge: 5-600)--making up the ground truth files in our dataset. Their input variants are described in \S\ref{sec:dataset-inputs}.

\subsubsection{10 Author Developed Programs} \label{sec:dataset-diy}
To test the LLMs on functions where data leakage is impossible, three authors manually crafted 10 new non-trivial single-file programs--writing both the C code and the VeriFast specifications--to be similar to programs in the 50-program dataset while not overlapping them. The resulting ground truth programs contain 45 functions, an ave. of 34 lines of code (rge: 17-55) and 32 lines of specifications (rge: 18-45). Six of the functions are concurrent, 3 contain loops, 20 require recursive predicates during verification, and 16 are other non-trivial SL dependent functions. Their input variants follow \S\ref{sec:dataset-inputs}. %

\subsubsection{Three Input Types} \label{sec:dataset-inputs}

\begin{table*}[t]
  \centering
  \includegraphics[width=0.85\linewidth]{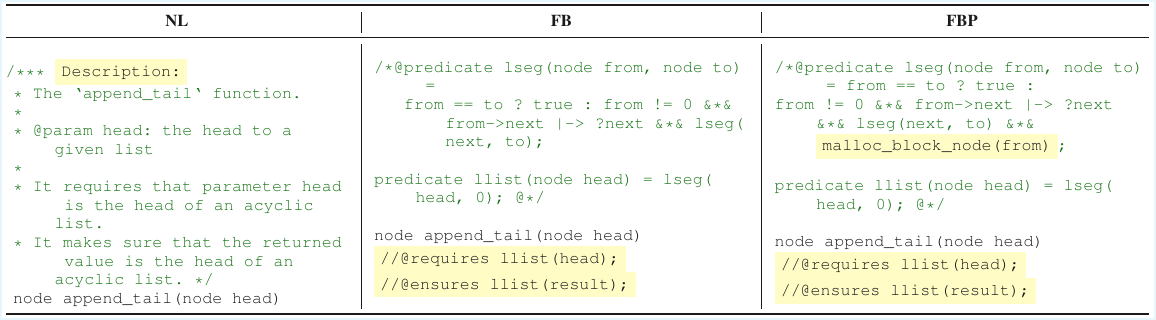}
   \caption{Example of 3 input types for \texttt{append\_tail} from \fig \ref{fig:ll-appendtail}, where the feature of each input type is highlighted in \lightyellow{yellow}.}
  \label{tab:input_type-examples}
\end{table*}

Inspired by Rego et al. \cite{rego2025evaluating}, for each of the 60 ground truth programs (303 functions), we developed three different, partially specified versions of them, which we refer to as Natural Language (NL), Functional Behavior (FB), and Functional Behavior Plus (FBP) versions:
\begin{itemize}
    \item[NL] The program contains only the source code and natural language description of its functional behavior, which is defined as the state of memory in logic before and after the function. For example, \tab \ref{tab:input_type-examples} shows the NL version of the \texttt{append\_tail} function on the left, which contains an informal description of the state of memory when entering and exiting the function in code comments (\eg ``the returned value is the head of an acyclic list'').
    \item[FB] The program contains source code and formal pre- and postconditions in VeriFast's specification language describing functional behavior. For example, \tab \ref{tab:input_type-examples} shows the FB format of \texttt{append\_tail} in the middle, where predicates \texttt{llist} and \texttt{lseg} capture the shape of an acyclic list for use in the pre- and postcondition. 
    \item[FBP] The program contains source code and formal pre- and postconditions that describe the program's functional behavior along with other specification constructs required to pass verification (\eg the allocation status of memory locations and constraints on integer bounds). For example, \tab \ref{tab:input_type-examples} shows the FBP format of \texttt{append\_tail} on the right, which additionally contains \texttt{malloc\_block\_node(from)} to aid in proving the absence of memory leaks in \texttt{append\_tail} and its callers. FBP is a baseline to compare NL and FB with.
\end{itemize}

Note, we ensure all input files for a function specify the same functional behavior as each other and the ground truth.

\subsection{Prompt Design and Selection with Pilot Study 1}
\label{sec:prompt_design_selection}

We explore various prompting methods that vary in content and input granularity similarly to Misu et al.~\cite{misu2024towards} and Du et al.~\cite{du2024evaluating}.
We consider a Basic (baseline), Chain-of-thought (CoT) \cite{wei2022chain}, RAG-sparse~\cite{bm25}, and RAG-dense~\cite{karpukhin2020dense}.
Our Basic prompt asks LLMs to generate verifiable specifications for a function given the input program and dependent standard VeriFast library files. The Basic prompt provides minimal guidance beyond requesting a single complete output code block.
The CoT prompt further instructs an LLM step by step through the specification process. Step 1 directs the LLM to write preconditions and postconditions for each function with sub-steps for functional behavior, memory safety, and integer bounds. It also enforces correct placement between the function declaration and body. Step 2 instructs the LLM to generate loop invariants where applicable, ensuring they hold at each iteration and entail the post-loop conditions. Finally, Step 3 asks the LLM to insert auxiliary inline statements, such as \texttt{open}, \texttt{close}, and lemma applications. %
Building on Basic, RAG-sparse and RAG-dense additionally provide the model with VeriFast tutorial text and programs chosen with sparse and dense retrieval accordingly.
Note, we do not include few-shot prompts, since RAG provides similar specified examples. %
We also consider 2 types of input granularity for each prompt type: input of the whole program (\textit{w/o splitting}), or one function from the program (\textit{w/ splitting}).
All prompts used are in the replication package described in \S\ref{sec:data-availability}.

\subsubsection{Pilot Study 1 for Prompt Selection} In total, we have 4 $\times$ 2 = 8 prompting methods. Since our analyses for RQ1 and RQ3 are primarily qualitative, it is infeasible to consider all 8 prompting methods across our entire dataset, inputs, and LLMs. Therefore, we conducted a pilot study with the 10 author-written programs' (45 functions') FBP inputs, GPT-4o, and our 8 prompts. We choose FBP since it is closest to a complete specification and GPT-4o due to its successes in related work \cite{janssen2024can,misu2024towards,rego2025evaluating}. The outputs are assessed for verification success rate, number of verification errors and their severity, and number of functional behavior changes in specifications and source code (defined in Section \S\ref{sec:QQA}). The results %
show RAG-sparse \textit{w/ splitting} has the highest verification success rate (64\% compared to at most 56\% for the others) and the fewest errors (63 compared to more than 67 for others). Moreover, it makes only 3 logical errors and 2 changes on functional behavior, meaning that LLMs are much less likely to make severe mistakes in the verification process or prove trivial results when using this prompting. Therefore, we selected RAG-sparse \textit{w/ splitting} to use in our full study.

\subsection{Pilot Study 2 for LLM Selection}
\label{sec:LLM_selection}
As with Pilot Study 1 (\S\ref{sec:prompt_design_selection}), it is infeasible to consider all 10 LLMs in our full study. As such, we conducted a similar pilot study using the selected RAG-sparse \textit{w/ splitting} prompt to choose the best three performing LLMs out of Claude 3.7 Sonnet (Claude-3-7), Deepseek-chat, Deepseek-reasoner, Gemini 2.5 Pro (Gemini-2.5), Llama 3.3 70B (Llama-3.3), Llama 4 Scout (Llama-4), Qwen3-32B (Qwen3), GPT-4o, GPT-4o-mini, and o3. %
We first checked the verification success rate. If the success rate is acceptable, then we examined the number of verification errors, their severity and the number of functional behavior changes on specifications and code. On this basis, %
we excluded Llama-3.3, Llama-4, Qwen3, and GPT-4o-mini due to a low success rate of $\leq 24\%$ (others are $\geq$ 40\%). Meanwhile, Claude-3-7, Gemini-2.5, and GPT-4o show the highest verification success rates at $\geq 60\%$ (the rest are $\leq 49\%$). Our qualitative analysis indicated they also have the fewest number of verification errors ($\leq 73$; others are $\geq 76$) and a small number of functional behavior changes ($\leq 2$), so these three LLMs are selected.

There is a potential bias in fixing GPT-4o for prompt selection and then using this selected prompt to select LLMs; 
so, we conducted an extra study to mitigate this issue. We explored the success rate of all 10 LLMs for each of the 8 prompts (80 combinations) across the 45 newly created functions' FBP inputs (since Claude-3-7 deprecated by the time of this study, we used Claude Haiku 4.5 as a close alternative). The chosen RAG-sparse w/ splitting prompt and RAG-sparse w/o splitting prompt are tied for the largest median number of verified functions (16; others are $\leq$ 14.5) across all LLMs. Further, the 3 chosen LLMs result in the largest median number of verified functions (23, 22.5, and 20.5; others are $\leq$ 15) across all prompts. Thus, the selection of our chosen prompt and LLMs are justified.

\subsection{Quantitative \& Qualitative Analyses}
\label{sec:QQA}

This section describes the quantitative and qualitative analyses we performed on the LLMs' output files to answer RQ1-3.
These analyses are performed per \textbf{function} rather than per \textbf{program} as in Rego et al.~\cite{rego2025evaluating} to produce more precise results.

\subsubsection{RQ1: How well do LLMs preserve functional behavior when generating specifications for proofs of correctness in VeriFast across different prompts, input types, and LLMs?}

To answer RQ1, we qualitatively analyzed whether a function's source code and pre- and postconditions from the input file are changed by the LLM in violation of the function's intended behavior. This ensures the LLM is not trying to prove a different and potentially trivial property compared to the input. For our analysis, at least two authors independently examined the corresponding input and output files and assigned an appropriate code. They reviewed each other's result and resolved any disagreement. Our methodology here, including codes, is inspired by Rego et al. \cite{rego2025evaluating}. %
For the output pre- and postconditions, we coded their functional behavior as ``equivalent'', ``strengthened'', ``weakened'', or ``other'' when compared to the input pre- and postconditions. Here, ``equivalent'' means the pre- and postconditions have the same semantics, and ``strengthened'' (``weakened'') means the output pre- and postconditions are semantically stronger (weaker) than the input ones.
The rest of the cases are coded with ``other'', meaning the semantics of the pre- and postconditions are changed.

For source code, we coded the outputs' functional behavior as ``unchanged'' or ``changed''. A function is marked as ``changed'' when it contains a compiler error or its source code semantics is changed from the input, \eg the output and input functions do not return the same values. %

\subsubsection{RQ2: How successful are LLMs at generating specifications that result in verified code with VeriFast across different prompts, input types, and LLMs?}

To answer RQ2, we ran VeriFast with default options (\eg checking integer overflow and heap chunk leakage) on the output functions with specifications and calculated the number and percentage of functions passing verification (\ie the \emph{verification success rate}).
This metric provides a succinct overview of specification quality and is widely used in prior works \cite{misu2024towards, ma2025specgen, mugnier2025laurel, yang2025autoverus}. %

\subsubsection{RQ3: When LLMs generate specifications that fail to verify with VeriFast, what errors are produced and how can they be fixed?}

\begin{table*}[t]
  \centering
  \includegraphics[width=0.85\linewidth]{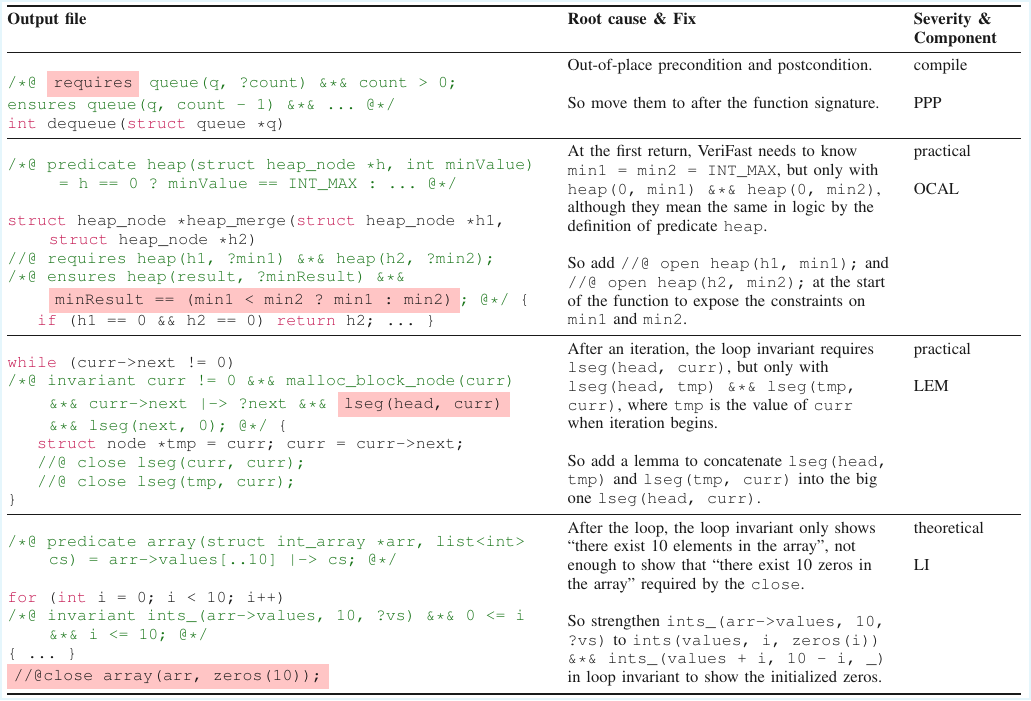}
  \caption{Examples of errors and analysis, where the errors in the output file are highlighted in \lightred{red}.}
  \label{tab:error-examples}
\end{table*}

To analyze errors produced by LLMs and the fixes, we sampled 50 error-prone functions out of 303, and examined all failing variants out of the 9 variants (across 3 LLMs and 3 input types) for each sampled function, totaling 354 functions. Our sampling process proceeded as follows. First, we calculated the verification success rates for each of the 303 functions across their 9 variants. Then, we grouped the functions by their language feature (concurrency, loops, recursive predicate, or other), decided on a success rate threshold for each feature category (33\%, 33\%, 44\%, 56\% respectively), and set a quota for each category (7, 7, 22, and 14) to match the dataset distribution. Finally, we select functions below the thresholds, first picking all the functions from the author dataset; and then, randomly sampling additional ones up to the quota for each category. %

For the analysis itself, we manually fixed failing output functions while recording errors, their type, location, and severity, and the corresponding fixes. VeriFast was used throughout this process to help locate and record errors and determine if the function is verified. This analysis is inspired by Rego et al. \cite{rego2025evaluating}, but we additionally track error severity. %
Two authors independently analyzed each output function and afterwards discussed and resolved conflicts in their coding. %
If the two authors fixed the function in similar ways but coded differently, then they adjusted the coding through compromise. %
But, if the two authors fixed the function in different ways, then we adopted the fix that %
has fewer modifications. %
Each error is tagged by severity represented by these codes: %
\begin{itemize}[leftmargin=0.7in]
    \item[Compile] This means VeriFast fails due to parsing or type check errors. %
    For example, the first error in \tab \ref{tab:error-examples} is caused by an incorrectly placed pre- and postcondition that will not parse correctly. %

    \item[Practical] This error is during the verification process. It is applied when a specification is insufficient for automated reasoning, but not logically incorrect in general nor logically insufficient for proof obligations.
    For example, the second error in \tab \ref{tab:error-examples} is ``practical'' since \texttt{open heap(0, min1)} is missing and VeriFast treats \texttt{heap(0, min1)} and \texttt{min1 == INT\_MAX} differently without this open even though they mean the same thing. %

    \item[Theoretical] This error occurs when verification fails due to logically incorrect or insufficient specifications, and is the most severe.  %
    For the last example in \tab \ref{tab:error-examples}, the invariant \texttt{ints\_(arr->values, 10, ?vs)} is too weak to imply \texttt{array(arr, zeros(10))}'s body required by the \texttt{close} after the loop. %
\end{itemize}

We also assigned each error a code corresponding to the specification component it pertains to:
\begin{itemize}[leftmargin=0.5in]
    \item[PPP] for predicate/fixpoint function/precondition/postcondition, specifying behavior and key proof obligations. %
    \item[OCAL] for open/close/assert/leak statements, which are simple auxiliary specifications. %
    \item[LEM] for lemma or built-in call, %
        which convert verifier information to a new form. %
    \item[LI] for loop invariant. %
    \item[SRC] for source code.
    \item[OTHER] for the rest of the cases (\eg ghost variables).
\end{itemize}

As examples, \tab \ref{tab:error-examples} additionally contains the modified component tag and descriptions of the fix for each error listed. %

\subsection{Experimental Setup}
To carry out the experiment on inputs, LLMs and prompts, we implemented a configurable script to interact with LLMs using their APIs, where we set the temperature as 0 to promote reproducibility. Our environment is a Yoga 14sACH 2021 machine with 16 CPUs of AMD Ryzen 7 5800HS Creator Edition, 16 GiB memory and 1 TB SSD, which runs Ubuntu 24.04 LTS and Python 3.12. For prompting with RAG, we used BM25 for sparse encoding and TextEmbedding for dense encoding (both from Python library fastembed), and stored the vectors on a Qdrant node with 0.5 vCPU, 1GiB memory and 4GiB disk. We used VeriFast 24-08-30 \footnote{\url{https://github.com/verifast/verifast/releases/tag/24.08.30}} to verify output files. %

\section{Results}
\label{sec:results}
For the full study, we fed all 3 input variants for each of our 303 functions and the RAG sparse \emph{w/ splitting} prompt into 3 LLMs resulting in 2,727 output functions. We analyzed the outputs according to \S\ref{sec:QQA} to answer RQ1-3 (stated in \S \ref{sec:study-design}).

\subsection{RQ1 - Functional Behavior}
\label{sec:FB}

\subsubsection{Source code}
\label{sec:FB-src_code}

\begin{table}[t]
\centering
\setlength{\tabcolsep}{4pt}
\begin{tabular}{l|ccc|c}
\hline
& Claude-3-7 & Gemini-2.5 & GPT-4o & \textbf{Total} \\
\hline
NL  & 21 & 24 & 19 & 64 (7.0\%) \\
FB  & 29 & 29 & 24 & 82 (9.0\%) \\
FBP & 36 & 45 & 19 & 100 (11.0\%) \\
\hline
\textbf{Total} & 86 (9.5\%) & 98 (10.8\%) & 62 (6.8\%) & 246 (9.0\%) \\
\hline
\end{tabular}

\caption{Number and percentage of changed functional behavior of source code in output files}
\label{tab:functional_behavior-src_code-result}
\end{table}

Table \ref{tab:functional_behavior-src_code-result} presents the number and percent of output functions that have their source code behavior changed for each LLM, input type, and overall. %
Only 246/2,727 (9.0\% of) output functions have their source code behavior changed, a promising result.
Among input types, NL shows the least amount of change with 64/909 functions (7.0\%) changed, while FBP shows the most with 100/909 (11.0\%) changed. For models, GPT-4o performs the best with 62/909 (6.8\%) changed; the other two models result in 86 and 98 functions (around 10\%) changed. The Cohen Kappa value \cite{viera2005understanding} of our coding here is 0.73, which is substantial agreement.

We further investigated the source code changes and present the results.  %
Out of the 246 changed functions, 102 (42\%) are marked as such due to small semantic edits, such as the addition/deletion of an if statement, change in variable usage, or giving library function declarations function bodies. We also observe edits leading to compiler errors, such as adding hallucinated header files or generating entire duplicate source code blocks, affecting 143/246 functions (58\%).

\begin{table}[t]
\centering
\setlength{\tabcolsep}{1.6pt}
\renewcommand{\arraystretch}{1.05}

\begin{tabular}{c|cccc}
\hline
 & eq & str & wk & oth \\
\hline
\textbf{Total} & 2375 (87\%) & 146 (5\%) & 116 (4\%) & 90 (3\%) \\
\hline
\end{tabular}

\vspace{4pt}

\begin{minipage}{0.48\columnwidth}
\centering
\begin{tabular}{l|ccccc}
\hline
 & eq & str & wk & oth & \%eq+str\\
\hline
NL  & 649 & 103 & 114 & 43 & 83\%\\
FB  & 859 & 25  & 2   & 23 & 97\%\\
FBP & 867 & 18  & 0   & 24 & 97\%\\
\hline
\end{tabular}
\end{minipage}
\hfill
\begin{minipage}{0.49\columnwidth}
\centering
\begin{tabular}{l|ccccc}
\hline
 & eq & str & wk & oth & \%eq+str\\
\hline
Claude-3-7 & 779 & 30 & 43 & 57 & 89\%\\
Gemini-2.5 & 802 & 72 & 17 & 18 & 96\%\\
GPT-4o     & 794 & 44 & 56 & 15 & 92\%\\
\hline
\end{tabular}
\end{minipage}

\caption{Distribution of functional behavior in precondition and postcondition in output files}
\label{tab:functional_behavior-PP-result}
\end{table}

\subsubsection{Preconditions and Postconditions}
LLMs do well at preserving specified functional behavior, as 2,375/2,727 (87\% of) output functions have pre- and postconditions (contracts) with equivalent behavior to the input specifications and an additional 146 (5\%) are strengthened in comparison (Table \ref{tab:functional_behavior-PP-result}). The NL input type performs noticeably worse than the FB and FBP types, but still preserves behavior in most cases. That is, for NL, 752/909 (83\% of) output contracts are equivalent or strengthened compared to the input vs. 884/909 (97\%) for FB and 885/909 (97\%) for FBP. Despite this difference, all 3 LLMs preserve specified behavior similarly well (Claude-3-7: 809/909 (89\%), Gemini-2.5: 874/909 (96\%), GPT-4o: 838/909 (92\%)). For this analysis, the Cohen Kappa is 0.38, \ie fair agreement. There were many conflicts (275+) when coding the NL outputs due to ambiguity in natural language. %

Upon further investigation, we see the NL input results in all but 2 weakened output contracts and nearly half (48\%) of all other (changed) contracts. This is not surprising as NL provides models with the least amount of formal specification guidance and leaves room for ambiguity. Notably, Gemini-2.5 contributes the least to these numbers, often choosing to strengthen output contracts instead when given more freedom. Lastly, 66/90 (73\% of) functions marked as having their specified behavior changed (\ie marked as ``other") are marked as such due to containing duplicate or missing specifications.

\subsubsection{Discussion}
Despite allowing LLMs to change source code and pre- and postconditions for verification, LLMs with prompting generally preserve the intended functional behavior of input source code and specifications. Users may also provide the input behavior of a function as formal specifications rather than natural language comments, and use Gemini-2.5 over other models to improve performance further. Gemini-2.5 is also notably robust to the ambiguity introduced by informal specifications, and is a good choice when formal specifications cannot be provided. %
However, LLMs cannot guarantee the absence of behavioral changes, and we witnessed a small percentage of such changes in our study. Some automated solutions exist to this problem, but come with trade-offs. First, LLMs can be instructed and/or forced to not modify the input code and specifications as with Banerjee et al. \cite{banerjee2026dafnypro} and Chen et al.'s \cite{chen2025automated} work. While sound, this approach severely limits the applicability of LLMs for verification (because strengthening specifications and modifying code are required to verify some programs).
A few recent work \cite{endres2024can,chen2025automated} use test cases to detect incorrect or weak specifications in LLM outputs, but their completeness depends on the availability and coverage of test cases. %
Therefore, mitigating, detecting, and addressing behavioral changes in code and specifications by LLMs is important future work for tool builders.

\begin{rqbox}
\textbf{Summary of RQ1:} LLMs with prompting generally preserve the input behavior of source code and specifications, even when specifications are provided informally. But, using formal input specifications and/or Gemini-2.5 (over other models) can boost performance.
While behavioral changes are rare (9.0\% for source code and 7\% for pre- and postconditions), they do happen and often result from adding/deleting if statements, changing variable usages, giving definitions to library function declarations, duplicating code and specifications, and hallucinating header files. Thus, achieving sound and complete guarantees of functional behavior preservation is an important open challenge. %
\end{rqbox}

\subsection{RQ2 - Verifiability}

\subsubsection{Overall Verifiability of LLM-Generated Specifications}
Across 2,727 generated output functions (303 functions $\times$ 3 input types $\times$ 3 LLMs), 855 verify successfully, yielding a 31.4\% verification success rate (about 1 in 3 functions). We explore this rate further with respect to input type and LLM (\S\ref{sec:verifiability-inputllm}) and functions' main language feature (\S\ref{sec:verifiability_by_language_feature}).

\subsubsection{Verifiability by Input Type and LLM}
\label{sec:verifiability-inputllm}
As shown in \tab \ref{tab:verifiability-result}, FBP has the highest verification rate at 39.9\% followed by FB at 29.5\% and NL at 24.6\%. Moreover, across all three LLMs, verification rates are highest with FBP inputs and generally lowest with NL inputs (with the exception of Claude-3-7). Therefore, stronger formal guidance improves outcomes, \ie supplying more complete input specifications leads to higher verification success.
Similarly, verification performance varies based on LLM choice. Overall, Gemini-2.5 outperforms Claude-3-7 and GPT-4o (with ave. rates 39.7\%, 29.8\%, and 24.5\% respectively). For each input type, Gemini-2.5 also achieves the highest success rates compared to the others.  %

\begin{table}[t]
\centering
\begin{tabular}{l|ccc|c}
\hline
 & Claude-3-7 & Gemini-2.5 & GPT-4o & \textbf{Average} \\
\hline
NL  & 26.1\% & 36.0\% & 11.9\% & 24.6\% \\
FB  & 24.1\% & 38.9\% & 25.4\% & 29.5\% \\
FBP & 39.3\% & 44.2\% & 36.3\% & 39.9\% \\
\hline
\textbf{Average} & 29.8\% & 39.7\% & 24.5\% & 31.4\% \\
\hline
\end{tabular}

\caption{Verification success rate by input type and LLM}
\label{tab:verifiability-result}
\end{table}

\subsubsection{Verifiability by Language Feature}
\label{sec:verifiability_by_language_feature}
Verification success rate varies substantially depending on a function's language features. Table~\ref{tab:verifiability-feature_vs_LLM} %
shows that Normal SL functions have an 51.2\% ave. success rate (much higher than the overall success rate of 31.4\%).
This suggests current LLMs can handle basic predicates and simple heap reasoning reasonably well. However, performance drops when recursive predicates are utilized (30.9\%), and falls even further for concurrent (12.3\%) functions and functions with loops (11.4\%). These types of functions require more complex reasoning (especially heap reasoning) with predicates, loop invariants, and lemmas.

We also find that Gemini-2.5 performs significantly better than Claude-3-7 and GPT-4o on concurrent functions and functions with loops (Gemini-2.5: 21.5\% and 21.4\%, Claude-3-7: 10.4\% and 9.4\%, and GPT-4o: 5.2\% and 3.4\%). Note, Claude-3-7 also outperforms GPT-4o by roughly 5-6\% on these functions.
This suggests LLM choice matters when verification requires non-trivial auxiliary proof structure. %

\begin{table}[t]
\centering
\small
\setlength{\tabcolsep}{3.5pt}
\resizebox{\columnwidth}{!}{%
\begin{tabular}{l|ccc|c}
\hline
 & Claude-3-7 & Gemini-2.5 & GPT-4o & \textbf{Average} \\
\hline
Normal SL & 51.0\% & 54.1\% & 48.6\% & 51.2\% \\
Recursive Predicate & 28.9\% & 42.0\% & 21.9\% & 30.9\% \\
Concurrency & 10.4\% & 21.5\% & 5.2\% & 12.3\% \\
Loop Invariant & 9.4\% & 21.4\% & 3.4\% & 11.4\% \\
\hline
\textbf{Model Gap (High-Low)} & 41.6 & 32.7 & 45.2 & 39.8 \\
\hline
\end{tabular}
}
\caption{Verification success rate by language feature and LLM}
\label{tab:verifiability-feature_vs_LLM}
\end{table}

\subsubsection{Discussion}
Prompting LLMs to generate verified functions with VeriFast achieves a modest 31.4\% verification success rate (about 1 in 3 functions). However, upon further investigation, we can suggest strategies to improve performance. First, users should provide intended behavior in formal specification form to reduce ambiguity and provide more guidance to the models. Second, the cost of running our full study for each LLM is: Gemini-2.5 at \$233, Claude-3-7 at \$232, and GPT-4o at \$171. While Gemini-2.5 is similarly priced to Claude-3-7, it significantly outperforms Claude-3-7, especially on harder functions containing concurrency and loops. GPT-4o has the lowest cost but also the worst performance. Considering, Gemini-2.5 also rarely changes the functional behavior of input programs, spending more on Gemini-2.5 for better overall performance may be worth it for verification tasks.
Finally, success is higher on the simpler end of our dataset (Normal SL functions verify at 51.2\%), but verification remains challenging for concurrent and loop containing programs (both near 12\%). Thus, tackling failures in cases that require substantial auxiliary proof structure will significantly improve performance.
We investigate these failure modes and provide actionable insights from them in \S\ref{sec:RQ3-result}.

\begin{rqbox}
\textbf{Summary of RQ2:} %
The selected prompt and LLMs show modest success at producing verified C code, and perform very poorly on tasks requiring more complex reasoning. Providing intended behavior as formal specifications can boost performance, as well as using Gemini-2.5 over other models. But, Gemini-2.5 has a higher cost, highlighting a trade-off between cost and performance.
\end{rqbox}

\subsection{RQ3 - Verification Errors}
\label{sec:RQ3-result}

From our deep error analysis, we find the total number of verification errors across all LLMs and input types is 1,652 (\tab \ref{tab:deep_analysis-result-aggregated}) from the 354 selected failing functions. We present the results of our investigation into their severity, location, and fix/root cause. We also contextualize and provide actionable insights from the results.

\subsubsection{Error Severity}
Regarding error severity (\tab\ref{tab:deep_analysis-result-aggregated}), there are more compile--syntax and type--(383, 23\%) and practical errors (1,167, 71\%) than theoretical errors (102, 6\%). Thus, prompt guided LLMs largely struggle to generate specifications that are compilable and sufficient for automated reasoning with VeriFast. Fortunately, they generate only a small fraction of specifications that are logically inconsistent. 
Upon further inspection of practical errors, 858 (51.9\% of all errors) arise from not satisfying a required heap chunk (\eg, predicate or points-to assertion) during verification. For example, LLMs sometimes fail to write opens, closes, or lemmas that capture how the heap changes throughout a function for the verifier. This leads to failed heap chunk checks due to a lack of ownership. These errors would arise regardless of SL verifier, since these specification patterns are inherent to this broader class of verifiers.
There are also VeriFast unique practical errors: 58 (3.5\% of all) errors related to missing or incorrectly written malloc blocks for leak checking. %

\subsubsection{Error Location and Fix/Root Cause}
Returning to \tab\ref{tab:deep_analysis-result-aggregated}, about half of all errors (767, 46\%) can be fixed by modifying simple auxiliary specifications open/close/assert/leak (OCAL). They are either missing (58\%), redundant (31\%), misplaced (7\%), or wrong (4\%) in the output files. Errors on predicate/fixpoint function/precondition/postcondition (PPP) specifications are also significant (381, 23\%), with 44\% caused by compiler violations (syntax, 22\% and type, 13\%), 56\% caused by verification failures, such as missing or incorrectly writing a
malloc block (14\%) or bounds check (14\%). For lemma errors (LEM) (331, 20\%), 62\% of them are due to missing a lemma call and 23\% of them are due to redundant lemma calls or definitions. For errors on loop invariants (LI) (61, 3.6\%), 20\% of them are caused by missing loop invariants, 25\% are caused by them being too weak, and 20\% are caused by them being too strong.  %

\subsubsection{Errors by Input Type and LLM}

We show the number of errors broken down by input type and severity on the left of \tab \ref{tab:deep_analysis-result-input_type_and_LLM}. NL results in the highest number of errors (688), followed by FB (518) and FBP (446). %
NL also leads to the majority (62\%) of all compiler errors and nearly half of all theoretical errors (47\%). This is caused by the lack of guidance natural language comments provide to LLMs on writing compilable specifications and the ambiguity in comments' intention. %
All input types lead to a large number of practical errors, but FBP results in the fewest (339 compared to 404 for NL and 424 for FB). This is unsurprising as FBP contains specifications that are closest to a verified solution. %

\tab \ref{tab:deep_analysis-result-input_type_and_LLM} also breaks down the number of errors by LLM and severity. Gemini-2.5 has the fewest overall errors (424 compared to 546 and 682), practical errors (290 compared to 369 and 508), and theoretical errors (9 compared to 37 and 56). It also matches the other models on compiler errors. A closer inspection reveals that the lower volume of practical errors is primarily driven by the OCAL category (129 compared to 205 and 322). The discrepancy in theoretical errors stems from the categories of PPP (0 compared to 11 and 11) and loop invariants (0 compared to 7 and 18). %

\subsubsection{Discussion} LLMs with prompting primarily struggle to generate specifications that are compilable and sufficient for automated reasoning, such as OCAL specifications. As before, a straightforward way to reduce these errors is to provide formal input specifications and use Gemini-2.5. To reduce compiler errors, a repair loop with VeriFast's error messages can be considered because such messages contain information useful for locating and fixing compiler errors. A simpler solution would filter out failing cases and re-prompt until compilation succeeds, which is reasonable since compiler errors have a lower occurrence.

For automated reasoning errors, particularly OCAL errors, we investigated VeriFast's auto-feature that can automatically generate open and close specifications in simple cases. %
We find that 39.4\% open and close errors can be fixed with this feature. So, a mixture of LLM-based and symbolic techniques for specification/proof generation can provide better results.
Additionally, since the vast majority of automate reasoning errors are related to insufficient reasoning about heap changes throughout a function, having VeriFast expose symbolic heap state information to LLMs in repair loops may also reduce such errors. Related, better debugging techniques and error messaging when verification failure occurs should be explored for SL verifiers to aid LLMs in generation/repair. Current error messaging for heap-related failures does not often correspond to or properly pin-point the root cause of failures.

\begin{rqbox}
\textbf{Summary of RQ3:} LLMs' errors are mainly on reasoning about SL verifier-specific knowledge (\eg syntax and heap reasoning), rather than on general logic. Using techniques on the LLM side (\eg, feedback loop with heap information) or in symbolic execution (\eg, heap chunk searching) may be helpful to reduce such errors. %
\end{rqbox}

\begin{table}[t]
\centering
\small
\setlength{\tabcolsep}{3pt}
\resizebox{\columnwidth}{!}{%
\begin{tabular}{lccccccc}
\hline
 & PPP & OCAL & LEM & LI & SRC & OTHER & \textbf{Total} \\
\hline
Compile error     & 167 & 73  & 57  & 14 & 31 & 41 & 383  \\
Practical error   & 192 & 656 & 270 & 22 & 18 & 9  & 1167 \\
Theoretical error & 22  & 38  & 4   & 25 & 6  & 7  & 102  \\
\hline
\textbf{Total}              & 381 & 767 & 331 & 61 & 55 & 57 & 1652 \\
\hline
\end{tabular}
}
\caption{Errors by severity and fix location in deep analysis}
\label{tab:deep_analysis-result-aggregated}
\end{table}

\begin{table}[t]
\centering
\resizebox{\columnwidth}{!}{
\begin{tabular}{lccc|ccc}
\hline
 & NL & FB & FBP & Claude-3-7 & Gemini-2.5 & GPT-4o \\
\hline
Compile & 236 & 66 & 81 & 140 & 125 & 118 \\
Practical & 404 & 424 & 339 & 369 & 290 & 508 \\
Theoretical & 48 & 28 & 26 & 37 & 9 & 56 \\
\hline
\textbf{Total} & 688 & 518 & 446 & 546 & 424 & 682 \\
\hline
\end{tabular}

}
\caption{Error severity by input type and LLM in deep analysis}
\label{tab:deep_analysis-result-input_type_and_LLM}
\end{table}

\section{Threats to Validity}
\label{sec:threats-to-validity}
\paragraph{Internal} %
A key internal threat stems from benchmark construction: inputs are derived from verified VeriFast programs, ensuring a well-scoped task but limiting scenarios where LLMs must also refactor and verify arbitrary source code.
The pilot studies to select the best prompt approach and best 3 LLMs may introduce selection bias. We show with a wider study without this bias that our selections are still the best choices (\ref{sec:LLM_selection}).
Reproducibility is another concern, %
as we analyzed only one output per function per configuration to control qualitative analysis cost. We set the temperature of the LLMs to 0 to reduce randomness.
Our qualitative analyses have the authors apply codes and fixes, which is a subjective process open to interpretation. %
Thus, we made sure at least two authors code independently and discuss until an agreement is reached. For RQ1, inter-rater agreement is moderate to substantial (Cohen Kappa is 0.73 and 0.38). For RQ3, we haven't found a proper metric due to its open-ended coding.
Finally, since the majority of our test dataset comes from a public repository, data leakage is possible. %
Our 10 new programs used for LLM and prompt selection and the full study have not been leaked. Plus, the relatively low verifiability on programs (31.4\%) suggests limited leakage effects.

\paragraph{External} %

External validity is limited by our choice of benchmarks, language, and verifier. Our dataset contains 60 programs and 303 functions, which may not capture the diversity of codebases in terms of size, libraries, and specification conventions. However, our dataset is representative of real-world code and is feature diverse (\eg concurrency). We evaluated on C programs only, although VeriFast also supports Java and Rust. This is standard in related work, so we leave studying other programming languages to future work. %
Finally, we only study proof generation for VeriFast and not any other SL verifiers (\eg Viper or Gillian). While similar, they still have different specification languages and degrees of automation from each other, which may affect results. %
However, errors related to heap reasoning are likely to generalize across SL verifiers, since reasoning about heap ownership and separation is fundamental to all SL systems.%

\section{Related Work}

Early work in specification generation for static verifiers is primarily symbolic \cite{ta2017automated, calcagno2009compositional, calcagno2011compositional, vogels2011annotation} or heuristic based \cite{furia2010inferring, flanagan2001houdini}. With improvements in transformer-based language models, there has been a surge in recent work exploring their synthesis capabilities.
The vast majority of this recent work is for non-SL verifiers, such as Frama-C \cite{kirchner2015frama},
OpenJML \cite{cok2011openjml}, Dafny \cite{leino2010dafny}, and Verus \cite{lattuada2023verus}, which have different specification patterns (proofs) compared to SL verifiers.

For Frama-C, AutoSpec \cite{wen2024enchanting} few-shot prompts an LLM to fill in specifications one-by-one on a given C program for verification with Frama-C; while VeCoGen \cite{sevenhuijsen2025vecogen} prompts an LLM to generate both C programs and specifications. 
Jan{\ss}en et al.~\cite{janssen2024can} ask ChatGPT and Kamath et al.~\cite{kamath2024leveraging} ask multiple LLMs to generate inductive loop invariants validated with Frama-C.
SpecGen \cite{ma2025specgen} uses OpenJML to iteratively guide an LLM to generate specifications for verifying Java programs.
For Dafny, Laurel \cite{mugnier2025laurel} generates helper assertions using LLMs with novel prompting techniques; Silva et al.~\cite{silva2025inferring} further refines fault localization and specification inference to generate multiple assertions; Banerjee et al.~\cite{banerjee2026dafnypro} adds predefined proof strategies to LLMs and post-processes the generated Dafny annotations; Pascoal et al.~\cite{pascoal2025automatic} combines GPT-4o and Claude 3.5 Sonnet to generate loop invariants; and Poesia et al.~\cite{poesia2024dafny} combines an LLM and search to add annotations iteratively to a Dafny method until it verifies. Misu et al.~\cite{misu2024towards} empirically explore how well GPT-4 and PaLM-2 synthesize verified Dafny methods with prompting. Our empirical evaluations are similar except for our deeper error analysis which checks severity and fixes.
Both Yao et al.~\cite{yao2023leveraging} and Chen et al.~\cite{chen2025automated} use GPT models to generate specifications backed by Verus. Yao et al.~\cite{yao2023leveraging} breaks the verification tasks into smaller ones for GPT-4 to complete and combines the results via static analysis. Chen et al.~\cite{chen2025automated} uses GPT-4o for self-evolving specification generation and fine-tunes a new LLM to take over this process. AutoVerus \cite{yang2025autoverus} uses multiple LLM agents to generate and refine different types of specifications. In contrast, we do not use LLM feedback loops, since we focus on the capabilities of off-the-shelf LLMs with prompt engineering. %

Rego et al.~\cite{rego2025evaluating} is the only work other than ours, which explores the efficacy of LLMs to generate specifications for SL verification. Our work is heavily inspired by theirs. We extend and refine their preliminary work to explore more LLMs, prompting approaches, and benchmarks. Further, all of our analyses operate at the more precise granularity of functions rather than whole programs, and our error analysis additionally records the severity of errors. We also offer actionable insights from our results, while Rego et al. \cite{rego2025evaluating} do not.

\section{Conclusion}
This paper provides a comprehensive insight into how LLMs perform when asked to generate specifications for C programs that can be verified by VeriFast. After narrowing down the best performing mainstream LLMs to three and the prompt approach out of eight to one, we analyzed the output of these three LLMs across 303 functions using RAG-sparse (\emph{w/ splitting}) with three different input types. Our results show that while LLMs do preserve functional behavior, their overall verification success is modest. Their performance improves when input specifications are stronger; and among the three models, Gemini 2.5 Pro consistently has the highest verification rates. Further qualitative analysis shows recurring error patterns related to heap reasoning. This highlights the current limitation of LLMs in producing precise specifications required by separation logic tools, rather than misunderstanding the program. Future work can leverage this knowledge to improve AI techniques and verifiers to support generating auxiliary specifications with LLMs.

\section{Data Availability}
\label{sec:data-availability}
The artifacts of this work, including the dataset, prompts, scripts for quantitative analysis and our qualitative analysis, are provided at \url{https://zenodo.org/records/19570523}.

\bibliographystyle{IEEEtran}
\bibliography{IEEEabrv,myabrv,references}
\end{document}